\documentclass[journal=jacsat,manuscript=article]{achemso}
\usepackage{colortbl} 
\usepackage{graphicx} 

\usepackage{chemformula} 
\usepackage[T1]{fontenc} 
\usepackage{textcomp}
\setkeys{acs}{doi = true}
\setkeys{acs}{maxauthors=25}
\usepackage{fix-cm}
\usepackage{amsmath}
\usepackage{amsfonts}
\usepackage{amssymb}
\usepackage[table]{xcolor}
\usepackage{graphicx}
\usepackage{geometry}
\author{Brenna C. Bierman}
\affiliation[University of Wisconsin-Madison Chemistry]
{Department of Chemistry, University of Wisconsin-Madison, Madison, Wisconsin 53706, United States}
\author{Gillian Nolan}
\affiliation[University of Illinois Urbana-Champaign MS\&E]
{The Grainger College of Engineering, Department of Materials Science and Engineering, University of Illinois Urbana-Champaign, Urbana, IL 61801, USA}
\author{Hongrui Ma}
\affiliation[University of Wisconsin-Madison E\&CE]
{Department of Electrical and Computer Engineering,  University of Wisconsin-Madison, Madison, WI 53706, United States}
\author{Ying Wang}
\affiliation[University of Wisconsin-Madison E\&CE]
{Department of Electrical and Computer Engineering,  University of Wisconsin-Madison, Madison, WI 53706, United States}
\alsoaffiliation[UW-Madison Physics]
{Department of Physics, University of Wisconsin-Madison, Madison, WI 53706, United States}
\alsoaffiliation[University of Wisconsin-Madison MS\&E]
{Department of Materials Science and Engineering,  University of Wisconsin-Madison, Madison, WI 53706, United States}
\author{Pinshane Huang}
\affiliation[University of Illinois Urbana-Champaign MS\&E]
{The Grainger College of Engineering, Department of Materials Science and Engineering, University of Illinois Urbana-Champaign, Urbana, IL 61801, USA}
\alsoaffiliation[MRL]
{Materials Research Laboratory, University of Illinois Urbana-Champaign, Urbana, IL 61801, USA}
\author{Daniel A. Rhodes}
\email{darhodes@wisc.edu}
\affiliation[University of Wisconsin-Madison MS\&E]
{Department of Materials Science and Engineering,  University of Wisconsin-Madison, Madison, WI 53706, United States}
\alsoaffiliation[UW-Madison Physics]
{Department of Physics, University of Wisconsin-Madison, Madison, WI 53706, United States}

\title[An \textsf{achemso} demo]
  {Superconducting Sn-Intercalated TaSe$_2$: Structural Diversity Obscured by Routine Characterization Techniques}
\abbreviations{TMD,vdW,CDW,PXRD,SCXRD,EDS,XPS,RRR}
\keywords{}

\begin{document}

\begin{tocentry}





    \includegraphics[width=1\linewidth]{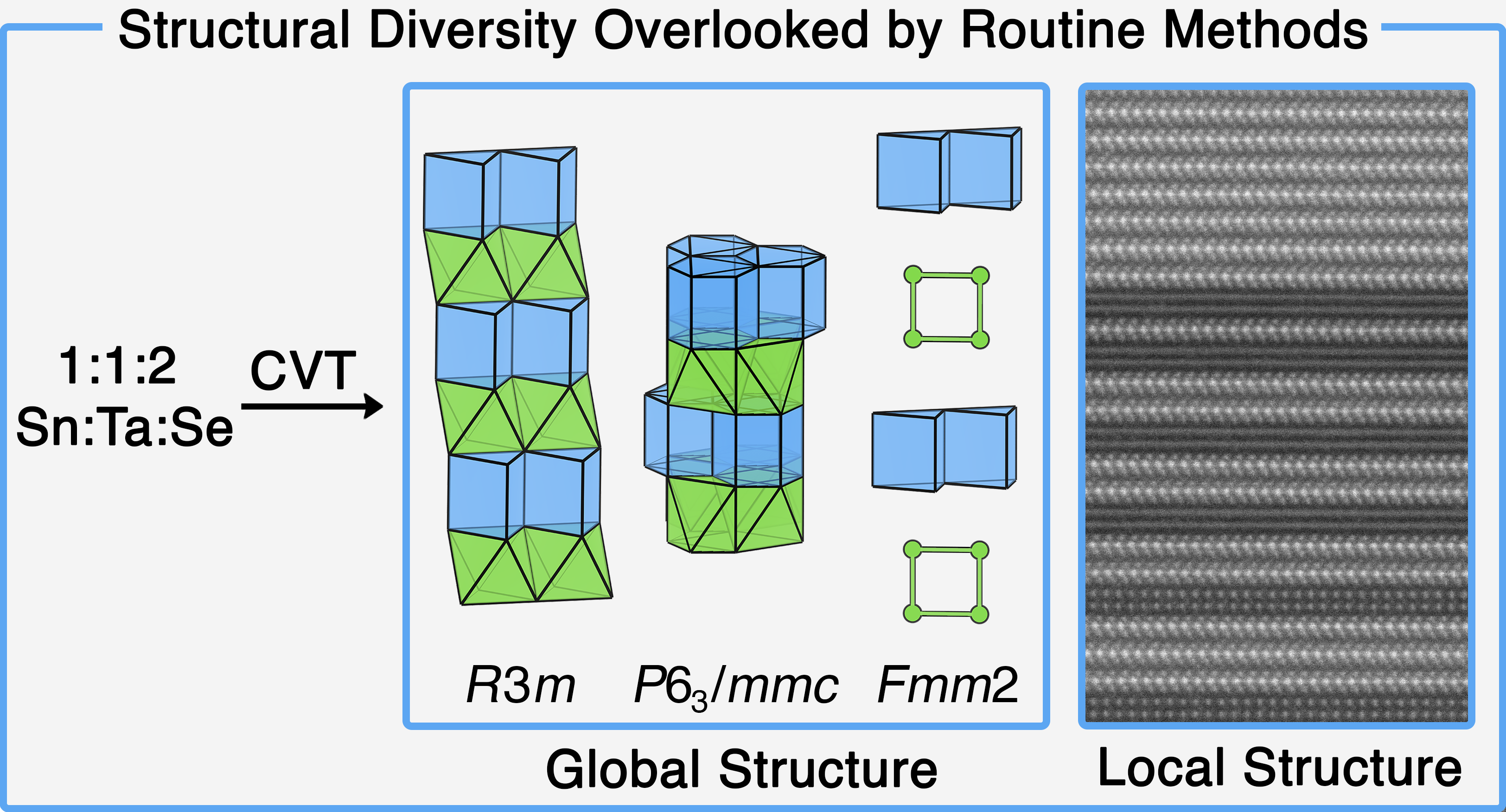}

 \end{tocentry}

\newpage
\begin{abstract}
 Using Sn-intercalated TaSe$_2$ as a model system, we demonstrate the presence of structural heterogeneity captured by single-crystal X-ray diffraction (SCXRD) and scanning transmission electron microscopy (STEM) that eludes the routine characterization techniques of powder X-ray diffraction, Raman spectroscopy, and electronic transport measurements. From a single growth composition (1:1:2 Sn:Ta:Se), we obtained crystals diverse in stoichiometry and structure, with near-continuous intercalation for Sn$_x$TaSe$_2$ from $0\lesssim{x}\lesssim1$. Using SCXRD, we found global structural diversity, identifying three new structure types: Sn$_{0.18}$TaSe$_{2.0}$/Sn$_{0.08}$TaSe$_{1.96}$ ($R3m$), Sn$_{0.16}$TaSe$_{2.0}$ ($P6_3/mmc$), and Sn$_{1.2}$TaSe$_{1.9}$ ($Fmm2$). Using STEM, we observed local structural diversity, manifested as regions of highly variable stacking within a single crystal. In contrast, powder X-ray diffraction did not resolve all observed global structures. Raman spectroscopy was unable to distinguish between different structures or compositions in the standard measurement range. Electronic transport measurements showed consistent superconductivity and charge density wave behavior irrespective of Sn-intercalation amount. Our results indicate that routine approaches to characterization of intercalated transition metal dichalcogenides may be inadequate for capturing the diversity of this family of materials, highlighting the need for high-resolution structural characterization when examining the properties of van der Waals-layered compounds.
\end{abstract}

\section{Introduction}
Transition metal dichalcogenides (TMDs) are a class of van der Waals (vdW) layered compounds known for their unusual electronic, optical, and physical properties, including superconductivity, charge density waves (CDW), and exfoliability~\cite{Manzeli2017}. Intercalation of TMDs with guest atoms can introduce or enhance properties such as magnetism~\cite{Jung2016,Nair2020} or superconductivity~\cite{Morosan2006,Zhang2016}. Intercalated atoms generally occupy an octahedral, tetrahedral, or trigonal prismatic void, forming distinct layers in the structures.~\cite{Whittingham1978,Kong2023} The intercalated atoms typically exhibit substantial positional and occupational disorder~\cite{Ali2015,Niu2024,Liu2025,Eder2025}. These non-periodic features can result in regions of local order in a crystal that may not be representative of the global crystal structure, complicating the determination of intercalated TMD structures.\cite{Li2019,Erodici2023,Goodge2023,Kochetkova2025} \\
\indent While metal-intercalated TMDs are well-studied~\cite{Rudorff1959,Whittingham1978}, much of the structural understanding of these compounds originates from characterization techniques that cannot resolve all structural details~\cite{Omloo1970,Eppinga1980}-- particularly powder X-ray diffraction (PXRD), which compresses three-dimensional diffraction patterns into one-dimension, resulting in information loss.~\cite{Harris1994,Flack1999,David2007,Leeman2024} 
Some TMD structures have been elucidated using more information-rich techniques, such as single-crystal X-ray diffraction (SCXRD)\cite{Meerschaut2001,Lai2021,Yang2021,Roseler2025}. However, many foundational reports on TMD structures employed now antiquated SCXRD methods, including determination of structures by hand from physical film.~\cite{Brown1965,Boebinger1983} SCXRD has since advanced, enabling the collection of more reflections and the determination of more accurate reflection intensities, but many structures have not been reexamined with modern techniques. As a result, reported structures may be incorrect or incomplete, potentially stymieing investigation of structure-property relationships.~\cite{Alnaser2023,Pollak2024,Devarakonda2024}\\ 
\indent For example, recent reevaluation of the structure of SrTa$_2$S$_5$ led to discovery of its unusual electronic characteristics~\cite{Devarakonda2024}. Early investigations of SrTa$_2$S$_5$ reported a $\sqrt{(28)}$a$\times\sqrt{(28)}$a supercell obtained using PXRD, but did not resolve a complete structure~\cite{Saeki1993,Kijima1996}. Devarakonda et al. reexamined and solved the complete crystal structure, clarifying the incommensurate modulation of SrTa$_2$S$_5$~\cite{Devarakonda2024}. This modulation was key to discovering the striped electronic behavior of SrTa$_2$S$_5$, suggesting this compound hosts a pair-density wave state.\\
\indent To highlight the potential complexity of and demonstrate an approach to more completely characterize intercalated TMD structures, we examined global and local structures of Sn-intercalated TaSe$_2$ crystals. SnTaSe$_2$ is predicted to be a topological superconductor candidate isostructural with PbTaSe$_2$~\cite{Chen2016}, a non-centrosymmetric superconductor known to host topological Dirac surface states~\cite{Guan2016,Chang2016}. To our knowledge, three studies of Sn-intercalated TaSe$_2$ have been detailed.~\cite{Driscoll1978,Gentile1979,Adam2021,Zheng2025} All reports rely primarily or exclusively on PXRD for global structural characterization. None of the reports used SCXRD or other long-range, high-resolution structural characterization methods. The earliest study reported a stoichiometry (SnTaSe$_2$) and lattice parameters (a = 3.42 \AA, c = 18.38 \AA) obtained by treating PXRD reflections as SCXRD reflections and constraining refinement using several assumptions about the structure, including the range of possible space groups~\cite{Driscoll1978,Gentile1979}. The stoichiometry was possibly checked with density measurements, but notably the SnTaSe$_2$ phase does not have Mössbauer data to support the structural assignment as provided for all other phases in the report. The second study reports a stoichiometry of Sn$_{0.5}$TaSe$_2$ that appears to result from a misinterpretation of energy dispersive spectroscopy (EDS) data~\cite{Adam2021}. The authors propose a structure for Sn$_{0.5}$TaSe$_2$ primarily by matching three $(00l)$ peaks to those of similar phases described in historic reports~\cite{Gentile1979,Eppinga1980,Dijkstra1989}. The most recent report provides a unit-cell sized-view of local order obtained by cross-sectional high-resolution scanning transmission electron microscopy (STEM). The proposed stoichiometry, 2Sn-2TaSe$_2$, is supported by both EDS and X-ray photoelectron spectroscopy (XPS) data. However, direct analysis of global structure by an information-rich technique was not conducted. Global structure was characterized mainly through comparison of select experimental PXRD peaks to those generated for a potential $R3m$ structure.~\cite{Zheng2025}. All three of these reports imply that the synthetic Sn-intercalated TaSe$_2$ products are uniform in stoichiometry and structure; we find this system to be significantly more complex. 

\indent Here, we use SCXRD to identify three new Sn$_x$TaSe$_2$ crystal structures ($0\lesssim\textrm{x}\lesssim1$), originating from a single growth composition (1:1:2, Sn:Ta:Se). While significant global structural diversity is represented among these phases, the full range of possible Sn$_x$TaSe$_2$ structures may not be captured by those reported here. Using STEM, we identified regions of spontaneous variable local stacking order within a single crystal of Sn$_x$TaSe$_2$, demonstrating appreciable local structural diversity.  We found that common characterization methods, such as PXRD, Raman spectroscopy, and electronic transport, were unable to capture the heterogeneity of the Sn$_x$TaSe$_2$ system. PXRD only resolved two of the three structures identified by SCXRD, and lacked the specificity needed to unambiguously characterize some structural details, such as stoichiometry, for the phases. Whereas stoichiometry and structure varied considerably, the Raman peaks in the standard measurement range, superconducting critical temperature ($T_\textrm{c}$), and charge density wave onset temperature ($T_\textrm{CDW}$) did not differ significantly across the samples measured. Our findings illustrate that a surprising amount of structural and stoichiometric diversity can emerge from a single growth composition and demonstrate the need for high-resolution approaches to intercalated TMD structure determination and verification.
\begin{figure} 
    \centering
    \includegraphics[width=.9\linewidth]{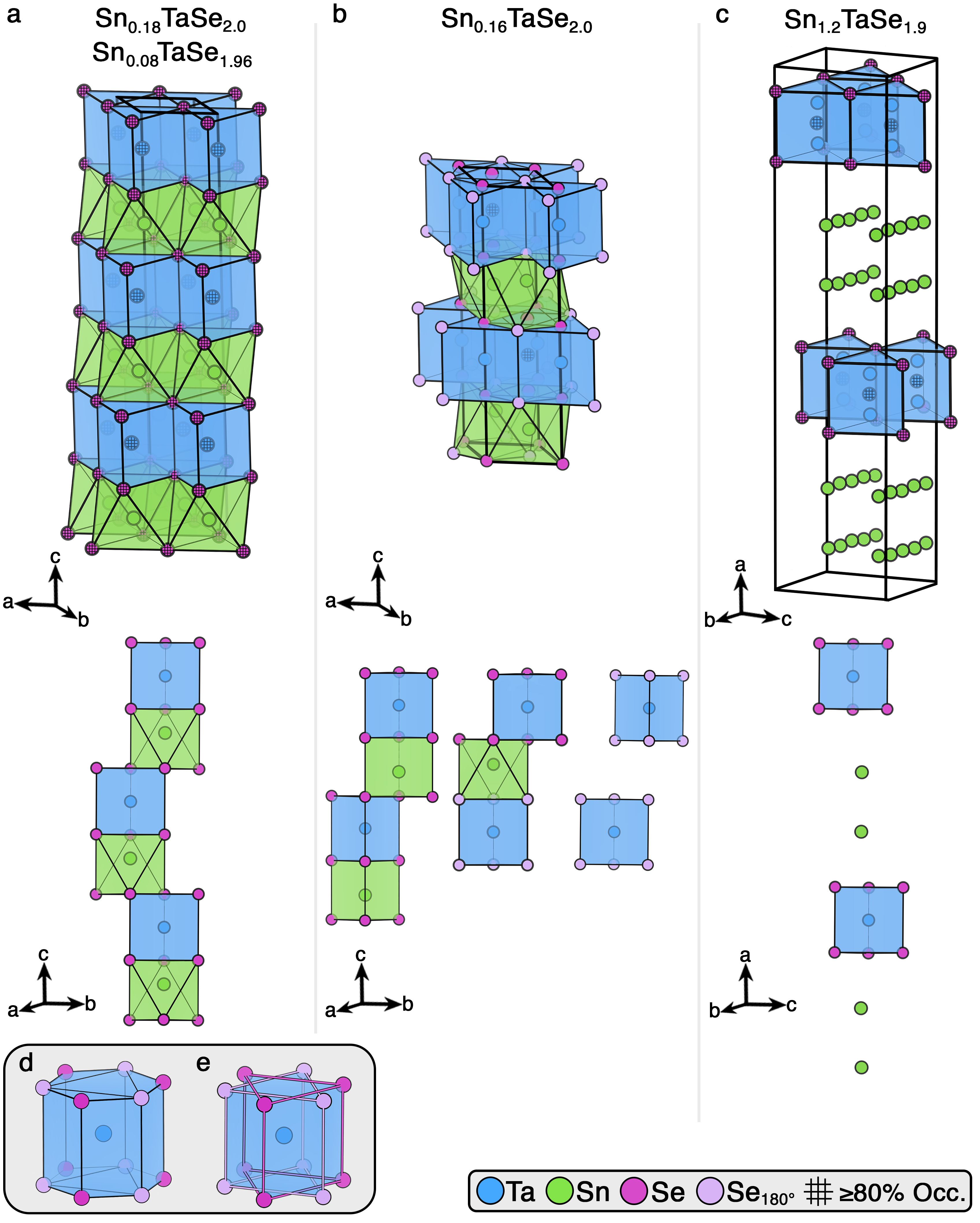}
    \caption{Crystal structures (top), idealized stacking (middle), and rotational disorder (bottom) for Sn$_\textrm{x}$TaSe$_2$. 
    (a) Sn$_{0.18}$TaSe$_{2.0}$ (structure one, dataset one) and Sn$_{0.08}$TaSe$_{1.96}$ (structure one, dataset two).
    (b) Sn$_{0.16}$TaSe$_{2.0}$ (structure two). 
    (c) Sn$_{1.2}$TaSe$_{1.9}$ (structure three). 
    (d) Ta coordination for Sn$_{0.16}$TaSe$_{2.0}$ as depicted in b. (e) Ta coordination for Sn$_{0.16}$TaSe$_{2.0}$ highlighting the two rotated trigonal prisms arrangements. $\ge$80\% occupied sites are designated with a grid pattern. Se$_{180^\circ}$ denotes the Se site attributed to a 180$^\circ$ rotation of the major Ta-centered coordination polyhedron.}
    \label{structure}
\end{figure}
\section{Results and discussion}
\subsection{Structural Characterization}
Sn$_{x}$TaSe$_2$ crystals were synthesized by chemical vapor transport (see Supporting Information for details) and analyzed using SCXRD. All crystals screened were twinned, including those that were visually pristine and a few microns in length (Figure S1). Screening $\gtrapprox 100$ crystals per dataset was required to find a specimen where the major twin component contributed $\geq90\%$ of the reflections. Non-merohedral twin domains could be a common, intrinsic feature of bulk vdW-layered TMDs, found in both intercalated and non-intercalated TMDs (Figure S2).\cite{Hart2023} Extrinsic factors, such as mechanical strain induced by cutting the crystals, could also introduce twin components.~\cite{McGuire2015,Jang2024,Siddique2024} Both mechanisms may contribute to the observed SCXRD behavior. Additionally, merohedral twinning might be common in Sn$_{x}$TaSe$_2$, occurring in two of the four SCXRD datasets. 
\\ 
\indent
Figure \ref{structure} presents the observed Sn$_x$TaSe$_2$ crystal structures, with refinement details provided in Table~\ref{celldata} and precession images in Figures S3-S6. The first structure was observed in two datasets, corresponding to two different stoichiometries, Sn$_{0.18}$TaSe$_{2.0}$ and Sn$_{0.08}$TaSe$_{1.96}$, in the same three-layer stacking configuration ($R3m$). The structure is analogous to 3$R$-Ta$_{1.11}$Se$_2$ ($R3m$)~\cite{Bjerkelund1967}. In Ta$_{1.11}$Se$_2$, the octahedral voids between the three layers of Ta-centered trigonal prisms are partially occupied by Ta (11\%). Similarly, in Sn$_{0.18}$TaSe$_{2.0}$ and Sn$_{0.08}$TaSe$_{1.96}$, the octahedral voids between TaSe$_2$ layers contain a Sn site occupied 18\% or 8\% respectively. Se occupies two distinct sites in both 3$R$-Sn$_x$TaSe$_2$ structures, forming the vertices of the trigonal prismatic Ta coordination polyhedra. All of the Ta and Se sites are fully or near fully occupied without positional disorder. Both datasets for the first structure exhibit merohedral twinning. Merohedral twin domains produce perfectly coinciding reflections, enabling determination of twin contributions from refinement.\cite{Herbst-Irmer1998,Herbst-Irmer2006} Dataset one features an inversion twin component: ($\bar1$ 0 0 / 0 $\bar1$ 0 / 0 0 $\bar1$), 49(17)\%. In dataset two, the twin component represents a -180$^\circ$ rotation around c$^*$: ($\bar1$ 0 0 / 1 1 0 / 0 0 $\bar1$), 25.2(7)\%.

\begin{figure}[ht]  
    \centering
    \includegraphics[width=.5\linewidth]{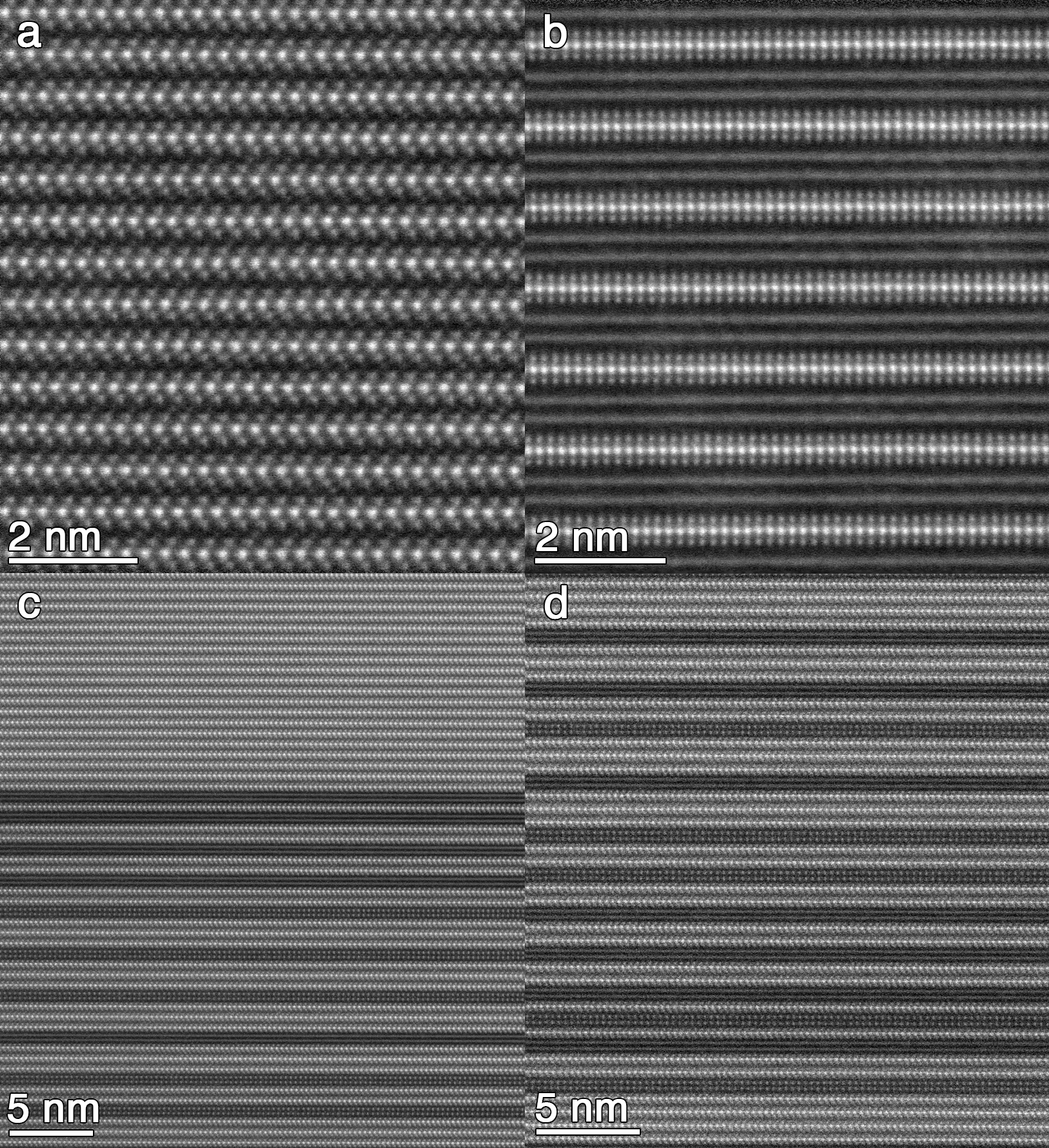}
    \caption{High-resolution cross-sectional STEM for Sn$_x$TaSe$_2$.  Periodic regions of (a) Ta-centered trigonal prisms with visible intercalants and
    (b) Ta-centered trigonal prisms alternated with two channels of Sn.
    (c) Transition from periodic to variable stacking.
    (d) Variability in number of TaSe$_2$ layers between Sn channels. STEM images for a, c, and d were obtained from the flake corresponding to structure two. STEM image for b was obtained from the crystal corresponding to structure three. All images show the out of plane direction as vertical.}
    \label{TEM}
\end{figure}

\begin{table}[t]
\centering
\caption{Structure Data and Refinement for Sn$_x$TaSe$_2$ Crystals}
\label{celldata}
\resizebox{\columnwidth}{!}{%
\begin{tabular}{>{\columncolor{gray!10}}c c c c c}
\hline
\rowcolor{white}
\textbf{Empirical Formula} & \textbf{Sn$_{0.18}$TaSe$_{2.0}$} & \textbf{Sn$_{0.08}$TaSe$_{1.96}$} & \textbf{Sn$_{0.16}$TaSe$_{2.0}$} & \textbf{Sn$_{1.2}$TaSe$_{1.9}$} \\ \hline
Radiation & Mo K$\alpha$ ($\lambda$=0.71069)  & Mo K$\alpha$ ($\lambda$=0.71069)  & Mo K$\alpha$ ($\lambda$=0.71069) &  Mo K$\alpha$ ($\lambda$=0.71069) \\ \hline
\textit{T}/K  & 293 & 293 & 293 & 293 \\ \hline
Crystal System  & Trigonal & Trigonal & Hexagonal & Orthorhombic \\ \hline
Space Group & \textit{R}3\textit{m} & \textit{R}3\textit{m} & $P6_3/mmc$ & \textit{Fmm}2 \\ \hline
a/Å  & 3.4417(1)  & 3.4409(8) & 3.44110(7) & 24.734(4) \\ \hline
b/Å  & 3.4417(1) & 3.4409(8) & 3.44110(7) &  3.4193(4) \\ \hline
c/Å  & 18.8554(9) & 18.997(5) & 12.622(1) &  6.0307(7) \\ \hline
Volume/Å$^3$ & 193.425(15) & 194.78(10) & 129.435(12) &  510.03(12) \\ \hline
Z & 3 &  3  & 2 &  4 \\ \hline
$\rho_\text{calc}$ g/cm$^3$ & 9.118 &  8.825  & 9.176 & 6.463 \\ \hline
$\mu$/mm$^{-1}$ &  70.853 & 70.161 &  71.554 &  42.678 \\ \hline
F(000) &  442.0 &  431.0  &  298.0 &  829.0 \\ \hline
Crystal Size/mm$^3$ & 0.03 × 0.02 × 0.01 & 0.01 × 0.007 × 0.001 & 0.0206 × 0.0133 × 0.001 & 0.0319 × 0.0188 × 0.0008 \\ \hline
2$\theta$ range for data collection/° & 6.482 to 57.16  & 6.434 to 58.932 & 6.456 to 59.988 & 3.294 to 51.564  \\ \hline
Index Ranges & 
\begin{tabular}[c]{@{}c@{}}-4 $\leq$ h $\leq$ 4\\ -3 $\leq$ k $\leq$ 4\\ -24 $\leq$ l $\leq$ 23\end{tabular} & 
\begin{tabular}[c]{@{}c@{}}-4 $\leq$ h $\leq$ 4\\ -4 $\leq$ k $\leq$ 4\\ -25 $\leq$ l $\leq$ 25\end{tabular} & 
\begin{tabular}[c]{@{}c@{}}-4 $\leq$ h $\leq$ 4\\ -4 $\leq$ k $\leq$ 4\\ -17 $\leq$ l $\leq$ 16\end{tabular} & 
\begin{tabular}[c]{@{}c@{}}-27 $\leq$ h $\leq$ 30\\ -4 $\leq$ k $\leq$ 4\\ -7 $\leq$ l $\leq$ 7\end{tabular} \\ \hline
Reflections collected & 1029 & 1249 & 1563 & 1122 \\ \hline
Independent reflections & 
\begin{tabular}[c]{@{}c@{}}168 [\textit{R}$_\text{int}$ = 0.0342,\\ \textit{R}$\sigma$ = 0.0247]\end{tabular} & 
\begin{tabular}[c]{@{}c@{}}184 [\textit{R}$_\text{int}$ = 0.0300,\\ \textit{R}$\sigma$ = 0.0194]\end{tabular} & 
\begin{tabular}[c]{@{}c@{}}100 [\textit{R}$_\text{int}$ = 0.0731,\\ \textit{R}$\sigma$ = 0.0277]\end{tabular} & 
\begin{tabular}[c]{@{}c@{}}296 [\textit{R}$_\text{int}$ = 0.0652,\\ \textit{R}$\sigma$ = 0.0576]\end{tabular} \\ \hline
Data/Restraints/Parameters & 168/0/14 & 184/0/14 & 100/1/19 & 296/0/12 \\ \hline
Goodness-of-fit on F$^2$ & 1.276 & 1.085 & 1.320 & 1.207 \\ \hline
Final R Indexes [I>2$\sigma$(I)] & 
\begin{tabular}[c]{@{}c@{}}\textit{R}$_1$ = 0.0266,\\ \textit{wR}$_2$ = 0.0686\end{tabular} & 
\begin{tabular}[c]{@{}c@{}}\textit{R}$_1$ = 0.0148,\\ \textit{wR}$_2$ = 0.0342\end{tabular} & 
\begin{tabular}[c]{@{}c@{}}\textit{R}$_1$ = 0.0441,\\ \textit{wR}$_2$ = 0.1076\end{tabular} & 
\begin{tabular}[c]{@{}c@{}}\textit{R}$_1$ = 0.0943,\\ \textit{wR}$_2$ = 0.2429\end{tabular} \\ \hline
Final R Indexes [all data] & 
\begin{tabular}[c]{@{}c@{}}\textit{R}$_1$ = 0.0266,\\ \textit{wR}$_2$ = 0.0686\end{tabular} & 
\begin{tabular}[c]{@{}c@{}}\textit{R}$_1$ = 0.0148,\\ \textit{wR}$_2$ = 0.0342\end{tabular} & 
\begin{tabular}[c]{@{}c@{}}\textit{R}$_1$ = 0.0469,\\ \textit{wR}$_2$ = 0.1090\end{tabular} & 
\begin{tabular}[c]{@{}c@{}}\textit{R}$_1$ = 0.1027,\\ \textit{wR}$_2$ = 0.2498\end{tabular} \\ \hline
Largest Diff. Peak/Hole/eÅ$^{-3}$ & 2.36/-1.57 & 0.67/-0.80 & 4.20/-5.40 & 4.89/-2.60 \\ \hline
Flack Parameter ($x$)  & 0.49(17) & -0.03(8) &  & 0.1(4) \\ \hline
Hooft Parameter ($y$)  & -0.01(4) & -0.07(3) &   & -0.11(6) \\  \hline
\end{tabular}
}
\end{table}
Structure two, Sn$_{0.16}$TaSe$_{2.0}$ ($P6_3/mmc$) (Figure \ref{structure}b), features a two-layer stacking arrangement of Ta-centered trigonal prisms, with interstitial Sn between TaSe$_2$ layers (8\%). Sn$_{0.16}$TaSe$_{2.0}$ appears to exhibit rotational disorder (Figure \ref{structure}d-e). Se is disordered across three sites. Two are \(\sim 50\%\) occupied, forming vertices for two trigonal prismatic arrangements around the major Ta site that appear 180$^\circ$ rotated around the $c$ axis, similar to the rotational disorder previously reported in 4$H$-Ta$_{1.03}$Se$_2$~\cite{Bai2018}. The disorder is not necessarily a 180$^\circ$ rotation. The observed sites could match a 60$^\circ$ rotation, 120$^\circ$ rotation, or a mirror plane. The rotational disorder creates three distinct interlayer environments (Figure \ref{structure}b idealized stacking). Two of these stacking arrangements can accommodate Sn, in trigonal prismatic or octahedral voids. The tetrahedral voids between two Se$_{180^\circ}$ layers are presumably vacant; if both Se$_{180^\circ}$ and Sn were present, the Se--Sn distance would be unreasonably short (1.2223(1) \AA) (Figure S7). Ta appears disordered across two sites (88\% and 11\%). The third Se site (13\%), or possibly the Se$_{180^\circ}$ site, appears to form a trigonal prismatic coordination environment around the second Ta site. Similar disordered positions have been reported before, and may be indicative of stacking faults~\cite{Luo2015,Cevallos2019}\\
\indent
The final structure, Sn$_{1.2}$TaSe$_{1.9}$ ($Fmm2$) (Figure \ref{structure}c), features the longest unit cell of the three structures, but contains only two TaSe$_2$ layers. The space between TaSe$_2$ layers contains two layers of two diffuse channels of Sn atoms, similar to the Sn layers reported in 2Sn-2TaSe$_2$\cite{Zheng2025}. The channels are composed of three distinct partially occupied Sn sites (15\%, 16.5\%, and 16\%). Positional disorder appears in Ta, with occupancy split between two sites (95\% and 5\%). Similar disorder was recently reported in Ba$_{0.75}$ClTaSe$_2$~\cite{Shi2024}, and might be explained by the presence of stacking faults.\cite{Takahashi1991,Shiojiri1993,Zhang2025} We note that our SCXRD analysis leaves some uncertainty in the structures. A discussion of this uncertainty is included in the Supporting Information. Elemental analysis provides additional support for the presence of Sn, in the form of XPS (Figure S8) and EDS (Figure S9-S10) data.\\ \indent
The three structures resolved by SCXRD represent distinct long range orders ($\mu$m scale) observed in the Sn$_x$TaSe$_2$ system. To examine the local ordering (nm scale), we collected high-resolution cross-sectional STEM. The samples analyzed by STEM were taken from the same bulk flakes as the specimens corresponding to structures two and three, but collection did not use the exact section of crystal analyzed by SCXRD. Figure \ref{TEM}a shows layers 
of Ta-centered trigonal prisms, with diffuse intercalants between layers. In Figure \ref{TEM}b, we observe layers of Ta-centered trigonal prisms alternating with two channels of Sn. In other regions, individual Sn atoms can be resolved (Figure S11). EDS of select regions were collected and support the presence of these features (Figures S12-S13). EDS did not definitively identify the intercalants shown in Figure \ref{TEM}a, but they could plausibly represent either Sn or Ta. The regions of periodic behavior demonstrate that consistent structures exist in the phase. We observe relatively consistent stacking, with expected local disorder represented in features such as stacking faults and inconsistent intercalant presence. 
\par
Our STEM data revealed a second dimension of structural diversity in Sn$_x$TaSe$_2$. Periodic stacking appears to spontaneously transition to variable stacking within a single crystal (Figure \ref{TEM}c). Within the variable stacking region, we observe between one to four TaSe$_2$ layers alternated with two layers of Sn channels (Figure \ref{TEM}d). In this instance, the variable stacking region spans approximately 150 nm (Figure S14), terminating in one $\mu$m of periodic stacking of two layers of TaSe$_2$ interleaved with two layers of Sn channels (Figure S15). This stacking order spans from the end of the variable stacking region to the surface of the crystal (Figure S16). A similar stacking order is observed in 2Sn-2TaSe$_2$, but is distinct in the relative orientation of TaSe$_2$ layers.\cite{Zheng2025} The formation of multiple stacking polymorphs within a sample is well established in layered materials, particularly those synthesized by deposition\cite{DeJong2018,Hadland2019,Huang2019,Feng2024,Hilse2025,Fender2025}. Our results expand upon these previous reports by evidencing disparate structural motifs within a single bulk-grown vdW crystal. Bulk properties measurements are typically ascribed to a single structural phase. That several distinct structures can coexist in a single crystal highlights the importance of considering both the local and global order when examining structure-property relationships for vdW-layered materials.  
\subsection{Routine Characterization Methods}
\begin{figure}[ht]
    \includegraphics[width=.5\linewidth]{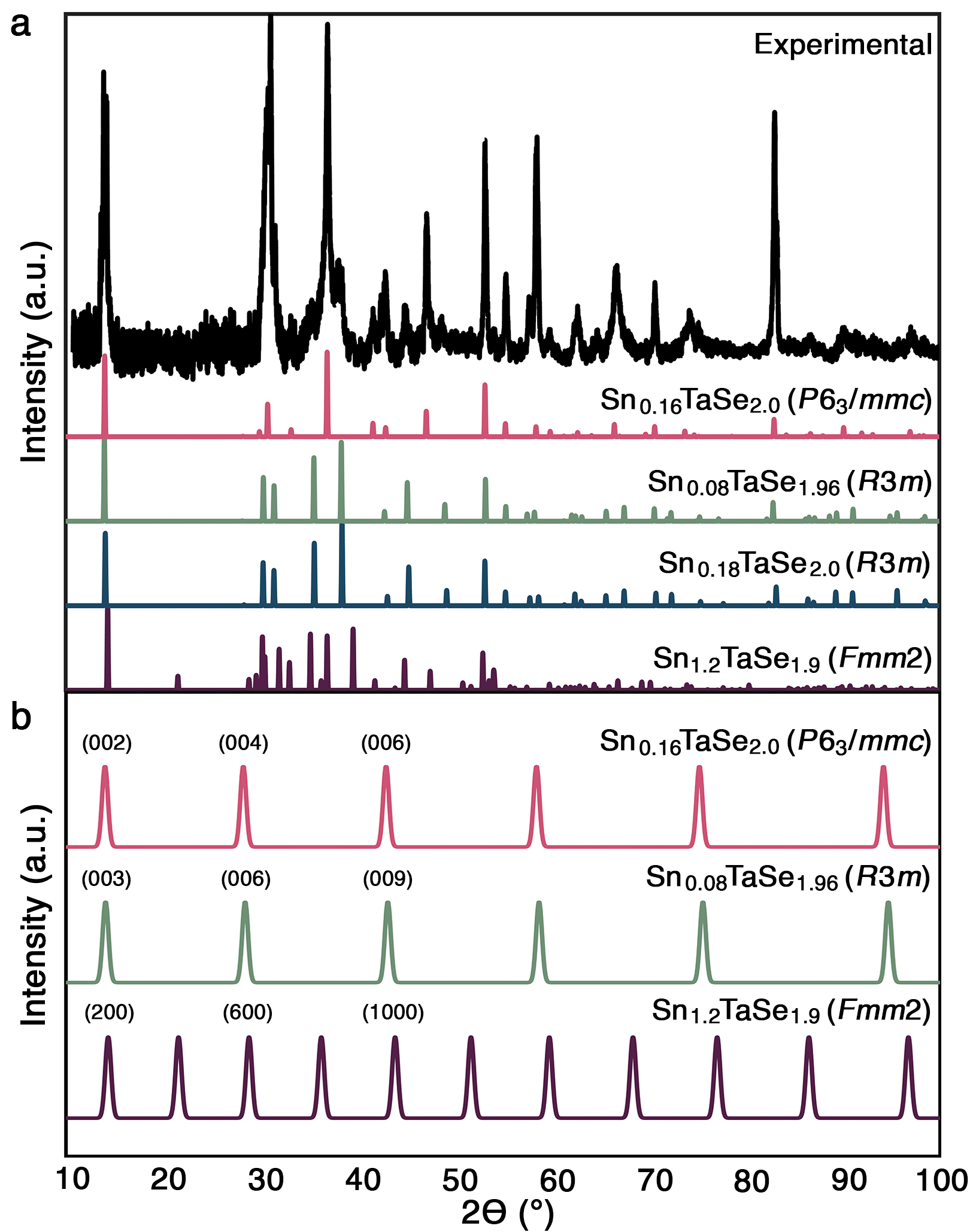}
   \caption{Sn$_x$TaSe$_2$ PXRD patterns. (a) Full experimental and theoretical patterns of the crystal structures observed by SCXRD and (b) ($00l$) (or analogous ($h00$)) peaks for the three Sn$_x$TaSe$_2$ structures with equalized peak intensities.}
  \label{PXRD}
\end{figure}

While we observed significant global and local structural diversity in Sn$_x$TaSe$_2$ by SCXRD and STEM, collected PXRD patterns appear to contain only two sets of peaks, likely matching structures one and two (Figure \ref{PXRD}a). The third structure was not identified. Unique peaks matching the orthorhombic structure are absent, potentially made indiscernible by a low signal to noise ratio. The weak signal intensity suggests the third structure is a minor phase. Reliance on PXRD alone would under-represent the structural diversity observable with more sensitive techniques, raising the possibility that minor phases in similar systems are being overlooked. 

\begin{figure}[ht]
    \centering
   \includegraphics[width=.5\linewidth]{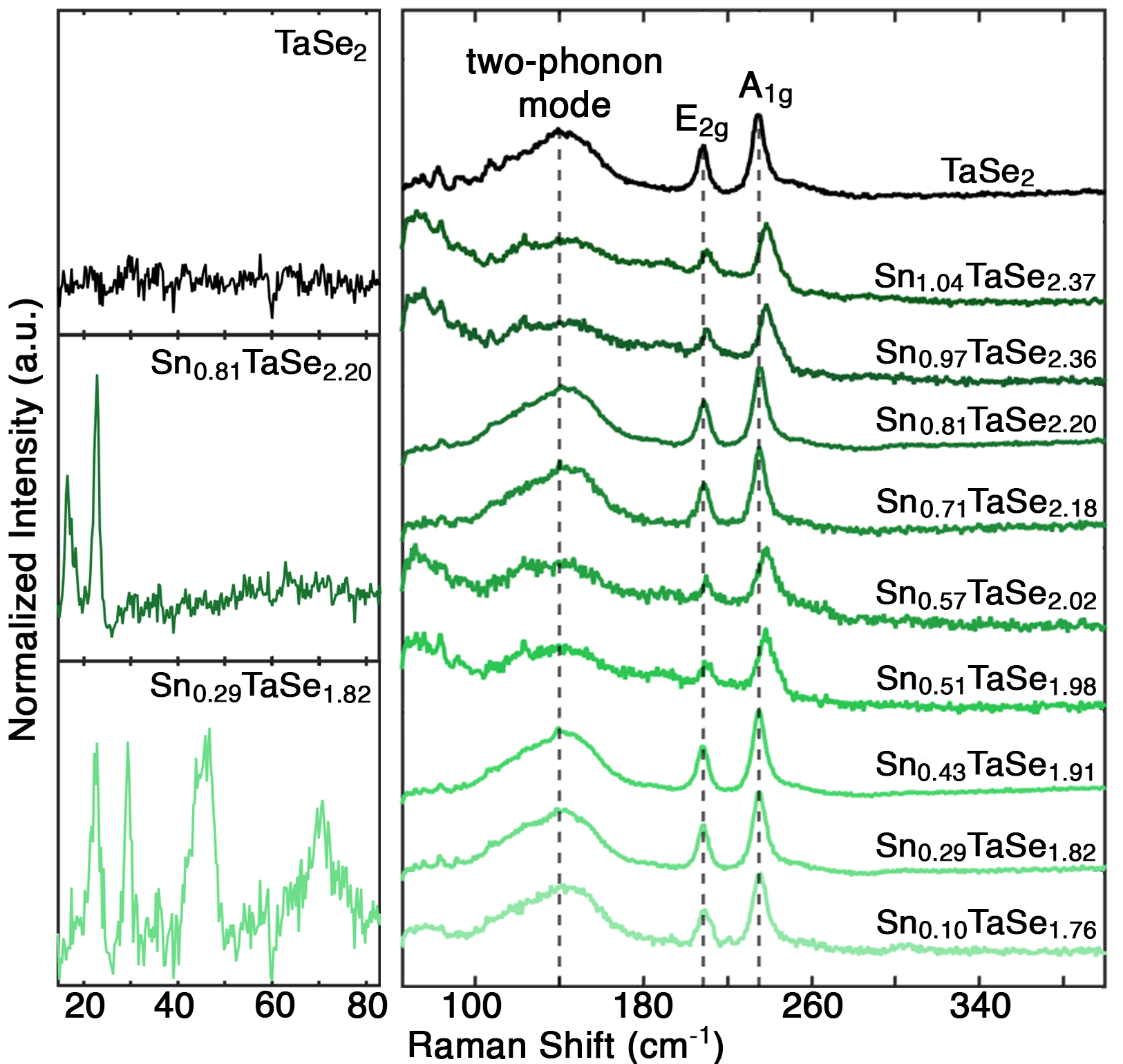}
   \caption{Raman spectra for Sn$_\textrm{x}$TaSe$_2$. Dashed lines indicate the position of the TaSe$_2$ peaks. Baseline correction and SiO$_2$ subtraction were used for low frequency data; raw spectra are presented in Figures S19-S22.}
  \label{Raman}
\end{figure}

\begin{figure}[ht]
   \centering
  \includegraphics[width=.5\linewidth]{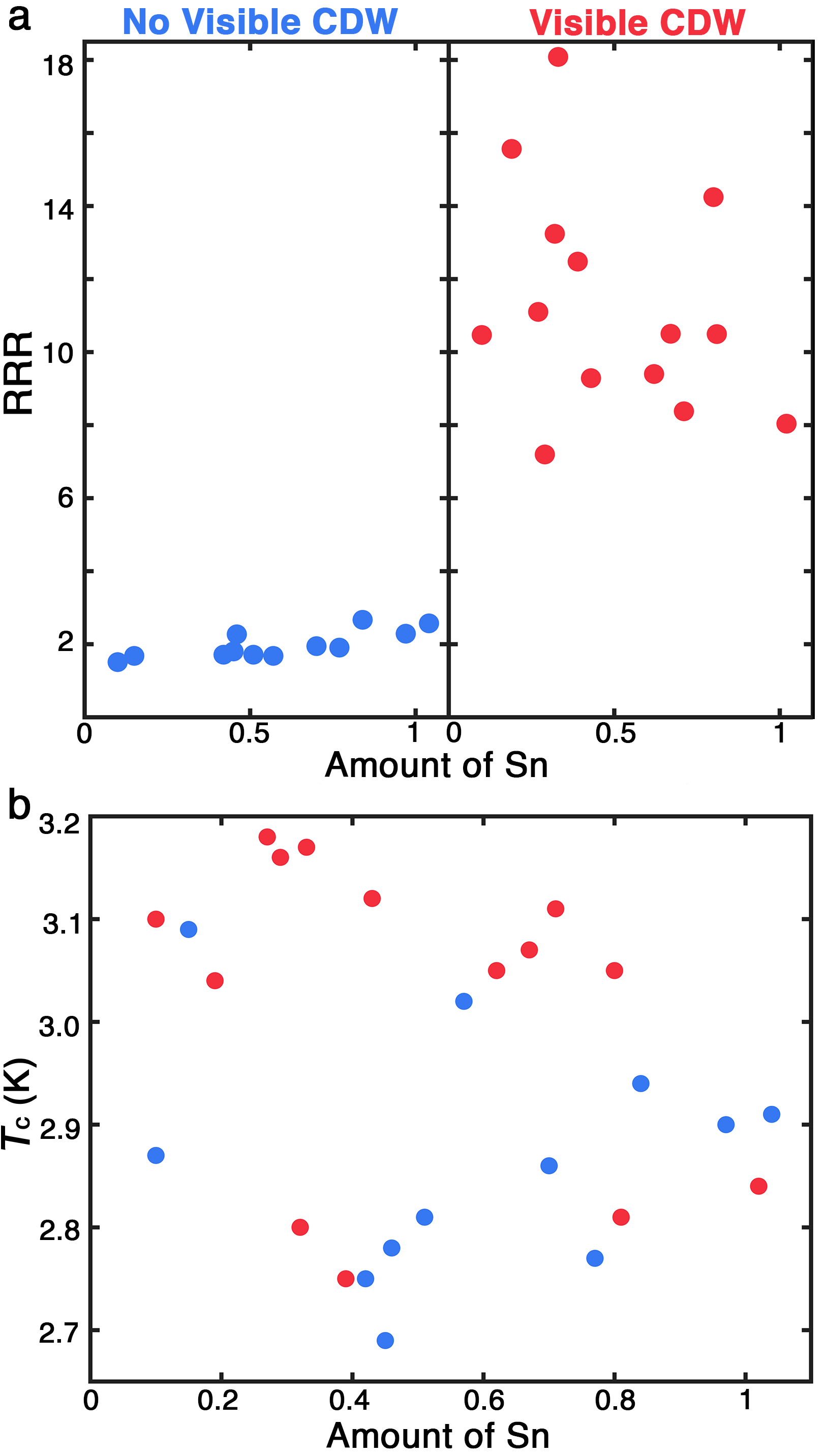}
   \caption{Transport behavior of Sn$_\textrm{x}$TaSe$_2$. (a) RRR plotted against the amount of Sn, partitioned according to the presence or absence of a visible CDW and (b) $T_c$ plotted against the Sn amount.}
    \label{Tc}
\end{figure}

Phase identification of layered materials like Sn$_x$TaSe$_2$ with PXRD is further complicated because significant changes to structure can result in only subtle changes to PXRD patterns. Stoichiometric changes, like those observed between the $R3m$ phases in Sn$_x$TaSe$_2$, can produce patterns that vary only by peak intensity, making differentiation challenging. Calculated PXRD patterns for structures with different space groups can also vary by peak intensities alone (Figure S17). Obtaining accurate intensities is challenging as layered 
structures often have a high degree of preferred orientation (Figure S18). Collecting a partial pattern further diminishes the sensitivity to structural changes. 
To avoid destructive powdering methods, phase identification of 
layered materials is often based on collection of $(00l)$ peaks of single crystals. The $(00l)$ peaks can provide insufficient specificity for structural determination even between structures with significantly different unit cells. Structures one and two have fully overlapping $(00l)$ peaks; structure three has peaks that overlap with those of structures one and two but has additional unique $(h00)$ peaks (Figure \ref{PXRD}b). Matching of $(00l)$ peaks would not allow for differentiation between structures one or two, and incomplete collection of peaks for structure three could result in mischaracterization of the phase. Our results demonstrate the limitations of PXRD as a tool for structural determination and verification, and underscore how reliance on this technique could result in overlooked structural diversity. 
\\
\indent
 In the  Sn$_{x}$TaSe$_2$ system we observed significant stoichiometric diversity in addition to the global and local structural diversity. Using EDS, we found a near-continuous range of Sn-intercalation between $0\lesssim{x}\lesssim1$. To understand the impact of stoichiometry on properties, we selected 26 single crystals representing the full range of Sn-intercalation for properties measurements (Table S1). 
 Given the significant structural differences observed by SCXRD and the variable stoichiometries measured by EDS, we expected the Raman spectra to vary significantly across the sample set~\cite{Tan2018,Fan2021,Erodici2023,Pereira2024}. However, the Raman spectra for all measured samples are nearly identical to pure $2H$-TaSe$_2$ for wavenumbers above 80 cm$^\textrm{-1}$, representing intralayer modes (Figure \ref{Raman})~\cite{Hajiyev2013,Neal2014,Hill2019}. Only small shifts in the A$_{1g}$ and E$_{2g}$ peaks were sporadically observed. These findings illustrate the limitations of Raman spectroscopy in verifying the structures of Sn$_\textrm{x}$TaSe$_2$ crystals, highlighting the need for careful multifaceted characterization approaches.
\\
\indent
Low-frequency Raman spectroscopy is known to distinguish between compounds with different symmetries by revealing interlayer breathing and shear modes that are sensitive to TMD stacking orders~\cite{Puretzky2015,Lee2016}. Figure \ref{Raman} shows the low-frequency Raman spectra obtained for Sn$_{x}$TaSe$_2$ samples representing no (x = 0), low (x = 0.29), and high (x = 0.81) Sn-intercalation. Spectral differences were observed between the three samples, with TaSe$_2$ having no observable low-frequency peaks, in agreement with previous ambient experimental results~\cite{Hill2019}; Sn$_{0.81}$TaSe$_{2.20}$ having two peaks at 16.5 and 22.8 cm$^{-1}$; and Sn$_{0.29}$TaSe$_{1.82}$ having four peaks at 22.8, 29.4, 45.8, and 70.8 cm$^{-1}$. These spectral differences imply that the compounds are distinct~\cite{Sam2020,Wang2022}. Some of the visible low-frequency peaks align with calculated inactive modes of 2$H$-TaSe$_2$, suggesting intercalation-induced symmetry breaking may have activated these modes~\cite{Hill2019,Chowdhury2021}.
Additional SCXRD and density functional theory phonon-calculations could connect the observed low-frequency modes with the structure to provide an avenue for optical phase verification that higher-frequency Raman spectroscopy cannot provide.\\
\indent Another common approach to verifying phase composition and structure involves measuring the electronic properties of a phase and matching measured to literature values. Small changes to structure are expected to produce clear changes in electronic behavior~\cite{Luo2015,Deng2020}. In Figure \ref{Tc}, we summarize our temperature-dependent resistivity ($\rho$) measurements for Sn$_x$TaSe$_2$ ($0.10\leq\textrm{x}\leq1.04$). For several crystals, we observe peaks in $\rho$ at $\sim111$~K, consistent with CDW peaks found in unintercalated TaSe$_2$. Despite significant variation in Sn, the $T_\textrm{CDW}$ does not significantly change across the sample set. The small variations in $T_\textrm{CDW}$ that are observed appear independent of stoichiometry (Figure S23). The absence of correlation between $T_\textrm{CDW}$ and stoichiometry is surprising, as $T_\textrm{CDW}$ typically changes with stacking order ($1T$-TaSe$_2$ $T_\textrm{CDW}$= 473~K (commensurate), 600~K (incommensurate)~\cite{Wilson1975,Craven1977}, $2H$-TaSe$_2$ $T_\textrm{CDW}$=90~K (commensurate), 120~K (incommensurate)~\cite{Kumakura1996}), and increased intercalation often suppresses CDW behavior~\cite{Wegner2019,Straquadine2019}.\\
\indent
As shown in Figure \ref{Tc}b, $T_\textrm{c}$ (defined as $0.9\rho_{3.3\textrm{K}}$) did not significantly vary with changes in stoichiometry. All measured Sn$_{x}$TaSe$_2$ crystals exhibited superconductivity, with only minor fluctuations in $T_\textrm{c}$ ($2.69\leq\mathit{T}_\textrm{c}\leq3.18$ K). Measurement of the magnetic susceptibility of a bulk sample corroborates this result (Figure S24). No clear trend or phase groups link $T_\textrm{c}$ to the amount of Sn or other changes in stoichiometry (Figures S25-S29). Magnetotransport results for Sn$_{x}$TaSe$_2$ displayed the same lack of correlation between stoichiometry and properties (Figure S30-S33). These results contrast with the observations of Luo et al., who found $T_\textrm{c}$ to be sensitive to intercalant amount and structure, particularly for the $3R$ phase of TaSe$_{2-x}$Te$_x$~\cite{Luo2015}.

\begin{figure}[ht]
  \includegraphics[width=.5\linewidth]{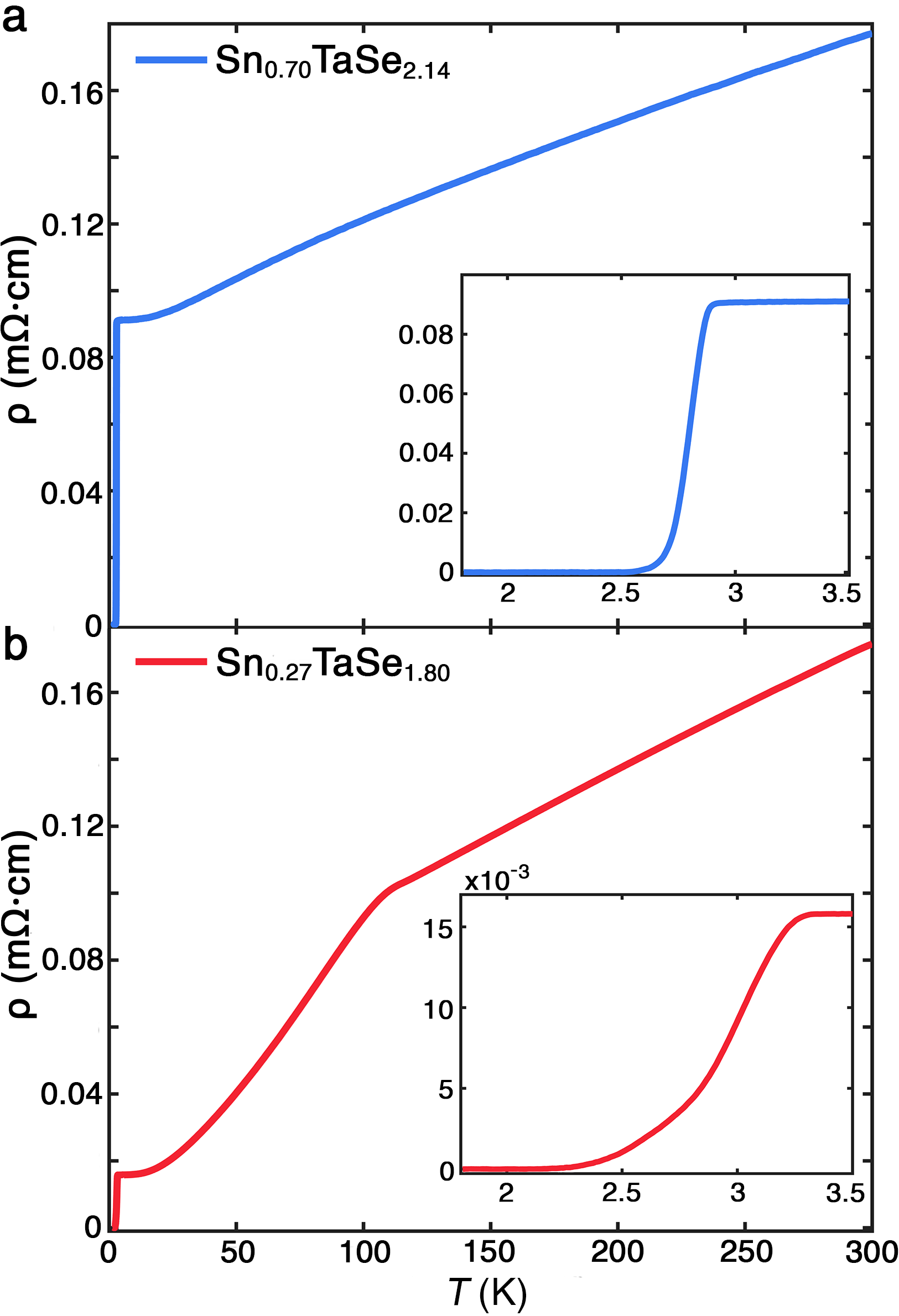}
  \caption{Representative Sn$_\textrm{x}$TaSe$_2$ transport curves. (a) Sn$_{0.70}$TaSe$_{2.14}$ (no visible CDW), inset shows the sharp superconducting transition. (b) Sn$_{0.27}$TaSe$_{1.80}$ (visible CDW), inset shows the broad superconducting transition.}
    \label{transport}
\end{figure}

\indent
As multiple phases were observed within a single crystal by STEM, the observed homogeneity in transport measurements 
could be a result of a single dominant phase or an averaging of all phases. Alternatively, all phases may exhibit comparable electronic behavior. In either case, the diversity of the system would be obscured if electronic properties were used as the metric for structure and composition verification. The observed similarity in $T_\textrm{CDW}$ and $T_\textrm{c}$ values could easily be mistaken for evidence of a consistent structure for Sn$_{x}$TaSe$_2$. \\ 
\indent
While $T_\textrm{c}$ does not correlate to changes in Sn$_{x}$TaSe$_2$ stoichiometry, the measured $T_\textrm{c}$ values are higher than those measured for pure TaSe$_2$ phases. The observed $T_\textrm{c}$ values represent a dramatic increase over that of $2H$-TaSe$_2$ ($T_\textrm{c}$ $\approx$ 0.2 K)~\cite{Kumakura1996}, and a slight increase over $3R$-TaSe$_2$ ($T_\textrm{c}$ $\approx$ 0.6 to 3.0~K)~\cite{Deng2020,Tanaka2020,Huang2023,Xie2025}. The substantial change in $T_\textrm{c}$ for TaSe$_2$ from $2H$ to $3R$ was attributed to changes in stacking order by Deng et al.~\cite{Deng2020}. However, fluctuations in $T_\textrm{c}$ linked to stacking order do not appear to be present in Sn$_{x}$TaSe$_2$. This seems to indicate that the origins of $T_\textrm{c}$ enhancement are not attributable to stacking order alone. Intercalation of Li or Pt into TaSe$_2$ has been reported to increase the $T_\textrm{c}$.\cite{Wu2019,LawanAdam2023}. The mutual presence of interstitial sites among $3R$-TaSe$_2$, Sn$_x$TaSe$_2$, and other intercalated TaSe$_2$ phases may point to involvement of these sites in the enhanced superconducting behavior.
\\ 
\indent Although electronic properties were insensitive to changes in stoichiometry, we observed a correlation between CDW behavior and crystal quality. Crystal quality, measured by residual-resistivity ratios (RRR), seems to control CDW appearance, with visible peaks present only in crystals with RRR $>$ 7 (Figure \ref{Tc}a). In addition, RRR appears positively correlated to $T_\textrm{CDW}$ (Figure S34). Correlation of $T_\textrm{CDW}$ with RRR has been observed in $1T$-VSe$_2$~\cite{Sayers2020}. Figure \ref{transport} presents $T$-dependent $\rho$ curves for samples with and without visible CDWs showing respectively broad or sharp superconducting transitions (for all measured flakes, see Figures S35-S60). 
\par
Similar broadening of the superconducting transition appears in CsV$_3$Sb$_5$ and is assigned to a nearly commensurate CDW state composed of multiple CDW domains~\cite{Yu2021}. The domain walls exhibit higher $T_\textrm{c}$, while the CDW domains exhibit varying degrees of suppressed superconductivity, generating inhomogeneous superconductivity. In Sn$_x$TaSe$_2$ crystals with visible CDW behavior, $T_\textrm{c}$ does not decrease as expected from the competition between superconducting and CDW phases~\cite{Yang2018,Wang2021a}. The absence of correlated variation may imply the CDW is present in all samples, but that flake quality controls its observability. Overall, the trends in electronic transport for Sn$_\textrm{x}$TaSe$_2$ appear to arise from changes in crystal quality, rather than structure or stoichiometry.

\section{Conclusion}
\indent Our identification of three new Sn$_x$TaSe$_2$ structures, representing a wide range of compositions and crystal symmetries, demonstrates that unexpected structural diversity can emerge from a single TMD growth composition. 
Complexity in structure and stoichiometry was identified between and within single crystals, illustrating both the global and local structural diversity found in the Sn$_x$TaSe$_2$ system. This heterogeneity was not reflected in PXRD, Raman spectroscopy, or electronic transport data. Materials parameters that were expected to be sensitive to changes in structure, like superconductivity, remained homogeneous across a wide range of stoichiometries. Investigations of intercalated TMDs often rely on assumptions that previously reported structures are correct and comprehensive, synthetic stoichiometries will match the stoichiometries of the products, and properties measurements can unambiguously verify the phase. Our findings suggest that these assumptions may not be valid. Given the intertwined nature of symmetry, topology, and superconductivity for this family of materials, accurate structures are necessary to understand their properties. We anticipate that our work will prompt others to characterize similar materials using multifaceted or high-resolution structural characterization techniques, potentially leading to the discovery of yet more structural diversity and unique physical phenomena.

\begin{acknowledgement}
 We are grateful to Dr. Ilia A. Guzei (UW–Madison) for assistance with SCXRD data collection and analysis and to Dr. Ashley Schmidt (Bruker AXS Applications Laboratory) for assistance with SCXRD analysis. We thank Philip P. Lampkin (UW-Madison) for fruitful discussions. This work was primarily supported by the National Science Foundation under grant No. DMR-2338984 and the University of Wisconsin–Madison. H.M. and Y.W. acknowledge support from the National Science Foundation through the University of Wisconsin Materials Research Science and Engineering Center (NSF DMR-2309000).  G.N. and P.H. acknowledge  the Materials Research Laboratory Central Facilities at the University of Illinois at Urbana-Champaign, and support by the Air Force Office of Scientific Research under award number FA9550-20-1-0302. The authors gratefully acknowledge the use of facilities and instrumentation in the Wisconsin Center for Nanoscale Technology. This center is partially supported by the Wisconsin Materials Research Science and Engineering Center (NSF DMR-2309000) and by the University of Wisconsin–Madison. The Bruker Quazar APEXII diffractometer used in this work was purchased by the University of Wisconsin-Madison Department of Chemistry with a portion of a generous gift from Paul J. and Margaret M. Bender. The Quantum Design MPMS3 magnetometer was supported by the University of Wisconsin--Madison Department of Chemistry.
\end{acknowledgement}

\begin{suppinfo}

The following files are available free of charge.
 \begin{itemize}
   \item SupportingInformation.pdf: Experimental and instrumentation details including additional discussion of structures, supplementary figures, and tables
 \end{itemize}

 CSD 2441345–2441348 contain the supplementary crystallographic data for this paper. These data can be obtained free of charge via www.ccdc.cam.ac.uk/data\_request/cif, or by emailing data\_request@ccdc.cam.ac.uk, or by contacting The Cambridge Crystallographic Data Centre, 12 Union Road, Cambridge CB2 1EZ, UK; fax: +44 1223 336033.
 \end{suppinfo}

\bibliography{References}

\end{document}